\documentstyle[aps,prl]{revtex}

\begin{document}

\title{Interference patterns of Bose-condensed gases in a two-dimensional
        optical lattice}

\author{Shujuan Liu$^{1}$, Hongwei Xiong$^{1}$, Zhijun Xu$^{1}$,
        and Guoxiang Huang$^{2}$}
\address{$^{1}$Department of Applied Physics, Zhejiang University of Technology,
               Hangzhou, 310032, China}
\address{$^{2}$Department of Physics and Key Laboratory for Optical and Magnetic
               Resonance Spectroscopy, East China Normal University, Shanghai
               200062, China}

\date{\today}
\maketitle

\begin{abstract}
{\it For the Bose-condensed gas confined in a magnetic trap and in a
two-dimensional optical lattice, the non-uniform distribution of atoms
in different lattice sites is considered based on Gross-Pitaevskii
equation. A propagator method is used to investigate the time
evolution of 2D interference patterns after (i)only the optical lattice is
swithed off, and (ii)both the optical lattice and the magnetic trap are
swithed off.  An analytical description on the motion of
side peaks in the interference patterns is presented by using the density
distribution in a momentum space.\\
{PACS number(s): 03.75.Fi, 05.30.Jp}}
\end{abstract}

\section{Introduction}

In the last few years, due to the experimental realization of
Bose-Einstein condensation in 1995 \cite{ALK}, we witnessed the remarkable
experimental and theoretical advances in the study of ultra-cold
Bose gases \cite{RMP}. Recently, a Bose-condensed gase confined in
an optical lattice (created by retroreflected laser beams) and in a magnetic
trap is investigated (see \cite{OPT} and references therein). In such system,
many subcondensates form and they are confined in the minimums of optical
lattice potential. This provides an ideal system for investigating
many interesting properties of Bose-Einsten condensate (BEC), such as
the quantum transition from a superfluid to a Mott insulating state of a
ultra-cold Bose gas \cite{GREINER}.

The coherence property of a BEC confined in an optical lattice and in a
magnetic trap can be obtained by observing the time evolution of
the interference patterns when both the optical lattice and the magnetic
trap are switched off. The interference patterns obtained in this way
is similar to the diffraction patterns of a coherent light from a grating.
For the BECs in one-dimensional (1D)\cite{PEDRI}, 2D \cite{2DOP} and 3D
\cite{GREINER} optical lattices, the interference patterns are clearly shown
experimentally. In \cite{GREINER}, the
quantum transition from a superfluid to a Mott insulating state is confirmed by
observing the characters of the interference patterns.

In a recent experiment, M\"{u}ller et al \cite{MORSCH} studied the time
evolution of the interference patterns when only a 1D optical lattice is
switched off. It is shown that,
due to the presence of the magnetic trap, there is a periodic harmonic motion
in side peaks.  This gives us a new opportunity to investigate the
properties of coherent matter wave such as the collision between the side peaks.
In our previous work \cite{XIONG}, a propagator method is used to study
the time evolution of the interference patterns formed in a 1D optical lattice,
and the harmonic motion of the side peaks predicted is in agreement with the
experimental result given in \cite{MORSCH}. In the
present work, we generalize the method developed in \cite{XIONG} to a 2D case,
i. e. we consider the time evolution of the interference
patterns for a Bose-condensed gas confined in a 2D optical lattice and in a magnetic
trap, with the
emphasis especially on the situation when only the optical lattice is
switched off.

\section{A Bose-condensed gas in combined trapping potentials}

For the Bose-condensed gas confined in a 2D optical lattice and
a magnetic trap with an axial symmetry along $z$-axis, the dynamics of the
condensate at low temperature is described by the well-known Gross-Pitaevskii
equation \cite{OPT}

\begin{equation}
i\hbar \frac{\partial \Psi }{\partial t}=\left[ -\frac{\hbar ^{2}}{2m}%
\bigtriangledown ^{2}+\frac{1}{2}m\omega _{\perp }^{2}(x^{2}+y^{2})+\frac{1}{%
2}m\omega _{z}^{2}z^{2}+V_{latt}+g\left| \Psi \right| ^{2}\right] \Psi,
\label{GP}
\end{equation}
where $\omega _{\perp }$ and $\omega _{z}$ are respectively the harmonic
angular frequencies of the magnetic trap in the radial and axial directions,
$g=4\pi^2 \hbar^2 a_s/m$ with $a_s$ the $s$-wave scattering length.
When the polarization vectors of two perpendicular optical standing waves
are orthogonal, the 2D optical lattice potential
takes the form \cite{2DOP}
\begin{equation}
V_{latt}=U_{0}\left[ \sin ^{2}\left( \frac{2\pi x}{\lambda }\right) +\sin
^{2}\left( \frac{2\pi y}{\lambda }\right) \right],  \label{latt}
\end{equation}
where $U_{0}$ denotes the depth of the optical lattice
which can be increased by increasing the intensity of retroreflected
laser beams, while
$\lambda$ is the wavelength of the retroreflected laser beams.
We consider here the case of $\omega _{\perp }<<\omega
_{z}$ and hence the condensate is disk-shaped with a larger extension
in the $x$ and the $y$ directions when in the presence of only the magnetic trap.
After the optical lattice potential $V_{latt}$ is switched on
a lot of subcondensates appear which are confined in the minimums (lattice sites)
of the optical lattice potential.
For the optical lattice created by the laser beams with wavelength
$\lambda$, $d=\lambda /2$ is the period of the optical lattice potential,
which can be regarded as the distance between two neighboring lattice sites.

We assume that the width of subcondensates in each lattice site, $\sigma$, is
much less $d$, which can be
realized in the present experiment by simply increasing the depth of the optical
lattice. We consider here the case that the subcondensates in different
lattice sites are fully coherent, i.e., the chemical potential of these
subcondensates are identical and hence the condensed state wave function (or called
order parameter) $\Psi $ of the
overall subcondensates can be written as:

\begin{equation}
\Psi =\sum_{k_{x},k_{y}}\Phi _{k_{x}k_{y}}\left( x,y,z\right) e^{-i\mu
t/\hbar },  \label{wavefunction}
\end{equation}
where $\mu $ is the chemical potential,
$k_{j}$ ($j=x,y$) denotes the $k_{j}$-th lattice site in the $j$-direction.

Substituting (\ref{wavefunction}) into Eq. (\ref{GP}), we have

\begin{equation}
\left[ -\frac{\hbar ^{2}}{2m}\bigtriangledown ^{2}+\frac{1}{2}m\omega
_{\perp }^{2}(x^{2}+y^{2})+\frac{1}{2}m\omega _{z}^{2}z^{2}+V_{latt}+g\left|
\Phi _{k_{x}k_{y}}\right| ^{2}\right] \Phi _{k_{x}k_{y}}=\mu \Phi
_{k_{x}k_{y}}.  \label{GP2}
\end{equation}
In obtaining Eq. (\ref{GP2}) we have used the condition that the overlap
between neighboring subcondensates can be neglected in the case of
$\sigma <<d$. To illustrate the character of $\Phi _{k_{x}k_{y}}$ clearly,
using the coordinate transformation
$x-k_{x}d\rightarrow x$, $y-k_{y}d\rightarrow y$
and making a Taylor expansion of the optical lattice potential to
the quadratic terms with respect to  the lattice
site $\{k_{x},k_{y}\}$, it is straightforward to get the following equation:

\[
{\ \left[ -\frac{\hbar ^{2}}{2m}\bigtriangledown ^{2}+\frac{1}{2}m\omega
_{\perp }^{2}(\left( x+k_{x}d\right) ^{2}+\left( y+k_{y}d\right) ^{2})+\frac{%
1}{2}m\omega _{z}^{2}z^{2}+\frac{1}{2}m\widetilde{\omega }_{\perp
}^{2}\left( x^{2}+y^{2}\right)\right. }
\]

\begin{equation}
\left. +g\left| \Phi _{k_{x}k_{y}}\right| ^{2}\right] \Phi _{k_{x}k_{y}}=\mu
\Phi _{k_{x}k_{y}},  \label{GP3}
\end{equation}
Equation (\ref{GP3}) determines the wavefunction $\Phi_{k_x k_y}$
at the lattice site $(k_x, k_y)$. In the
present experiment, the intensity of the retroreflected laser can be increased
so that one can have $\widetilde{\omega }_{\perp }>>\omega _{\perp }$.
For the subcondensate
confined in the lattice sites, one has $-d/2<x<d/2$ and $-d/2<y<d/2$. Thus
we obtain  $x<<k_{x}d$ and $y<<k_{y}d$ except for the subcondensate in the central
lattice site. As a result the term $\frac{1}{2}m\omega _{\perp
}^{2}(\left( x+k_{x}d\right) ^{2}+\left( y+k_{y}d\right) ^{2})$ in Eq. (\ref
{GP3}) can be approximated as $\frac{1}{2}m\omega _{\perp
}^{2}(k_{x}^{2}d^{2}+k_{y}^{2}d^{2})$. As for the
subcondensate in the central lattice site, because $\widetilde{\omega
}_{\perp }>>\omega _{\perp }$, the above  approximation is valid too. Thus
Eq. (\ref{GP3}) can be reduced to

\begin{equation}
\left[ -\frac{\hbar ^{2}}{2m}\bigtriangledown ^{2}+\frac{1}{2}m\omega
_{z}^{2}z^{2}+\frac{1}{2}m\widetilde{\omega }_{\perp }^{2}\left(
x^{2}+y^{2}\right) +g\left| \Phi _{k_{x}k_{y}}\right| ^{2}\right] \Phi
_{k_{x}k_{y}}=\mu _{k_{x}k_{y}}\Phi _{k_{x}k_{y}},  \label{GP4}
\end{equation}
where $\mu _{k_{x}k_{y}}=\mu -\frac{1}{2}m\omega _{\perp }^{2}d^{2}\left(
k_{x}^{2}+k_{y}^{2}\right) $,
which can be regarded as an effective chemical potential of the subcondensate
on the lattice site $\{k_{x},k_{y}\}$.
Assuming that
$\mu =\frac{1}{2}m\omega _{\perp }^{2}d^{2}k_{M}^{2}$, the effective
chemical potential then reads

\begin{equation}
\mu _{k_{x}k_{y}}=\frac{1}{2}m\omega _{\perp
}^{2}d^{2}(k_{M}^{2}-k_{x}^{2}-k_{y}^{2}).  \label{effective}
\end{equation}
Because the occupied lattice sites lie within a circle, $\pi k_{M}^{2}$ can
be regarded as the number of subcondensates induced by the optical lattice.

In the present work, we are interested in the case
$\hbar \omega _{z}<<\mu <<\hbar \widetilde{\omega }_{\perp }$. In this
situation, each subcondensate in the optical lattice is a cigar-shaped one along
the $z$-direction thus it displays a quasi-1D character. So a
Thomas-Fermi approximation can be applied in the $z$ direction.
After integrating out the fast-varying variables $x$ and $y$ in Eq. (\ref{GP4}),
one obtains \cite{PETROV}

\begin{equation}
\left[ -\frac{\hbar ^{2}}{2m}\frac{d^{2}}{dz^{2}}+\frac{1}{2}m\omega
_{z}^{2}z^{2}+g_{1D}\left| \varphi _{k_{x}k_{y}}\left( z\right) \right|
^{2}\right] \varphi _{k_{x}k_{y}}\left( z\right) =\mu _{k_{x}k_{y}}\varphi
_{k_{x}k_{y}}\left( z\right) ,  \label{quasi1D}
\end{equation}
where $\varphi _{k_{x}k_{y}}\left( z\right) $ is the wave function of the
subcondensate at the lattice site $(k_x,k_y)$ in the $z$-direction and
$g_{1D}=2\hbar^{2}a_s/m\widetilde{l}_{\bot }^{2}$ ($\widetilde{l}_{\bot }=\left( \hbar /m\widetilde{\omega }_{\perp }\right) ^{1/2}$)
is an effective coupling
constant in one dimension.

In the $z$ direction, using the Thomas-Fermi
approximation we obtain the atomic number $N_{k_{x}k_{y}}$ in the lattice site
$\{k_{x},k_{y}\}$ as

\begin{equation}
N_{k_{x}k_{y}}=\frac{4\sqrt{2}\mu _{k_{x}k_{y}}^{3/2}}{3g_{1D}m^{1/2}\omega
_{z}}. \label{Nkxky}
\end{equation}
Thus the ratio between the atomic number in the lattice site $(k_x,k_y)$ and the
atomic number in the central lattice site $(0,0)$ reads
\begin{equation}
N_{k_{x}k_{y}}=N_{00}\left( 1-\frac{k_{x}^{2}
+k_{y}^{2}}{k_{M}^{2}}\right) ^{3/2}.  \label{ratio}
\end{equation}
We see that the uniform chemical potential of the fully coherent
subcondensates and the presence of the magnetic trap result in a non-uniform
distribution of the atomic number in different lattice sites. It is worth
pointing out that the atoms can move from one lattice site to another ones
when the subcondensates are fully coherent (i.e. the system is in a
superfluid state). Thus the atomic number in lattice sites discussed here
should be regarded as an average one. In addition, the value of
$k_{M}$ can be obtained by considering the condition $N_{0}=%
\sum_{k_{x},k_{y}}^{^{\prime }}N_{k_{x}k_{y}}$. Here $N_{0}$ is the total
atomic number in all subcondensates, the prime in the sum
means that the summation about $k_{x}$, $k_{y}$ should satisfy the
condition $k_{x}^{2}+k_{y}^{2}\leq k_{M}^{2}$. Based on this
analysis, $k_{M}$ is given by

\begin{equation}
k_{M}=\left( \frac{15N_{0}g_{1D}\omega _{z}}{4\pi m\omega _{\perp }^{3}d^{3}}%
\right) ^{1/5}.  \label{km}
\end{equation}

In the present work, the parameters used are $\omega _{\perp }=24\times 2\pi$
Hz, $\omega _{z}=220\times 2\pi$ Hz and
$\widetilde{\omega }_{\perp }=10^{4}\times 2\pi $ Hz
for $N_{0}=10^{5}$  $^{87}$Rb atoms. Note that
$\widetilde{\omega }_{\perp }$ here is about one half of
the experimental value  used in \cite{2DOP}, thus the subcondensates discussed
here are fully coherent. It is obvious that
the condition $\hbar \omega _{z}<<\mu <<\hbar \widetilde{\omega }%
_{\perp }$ used above holds for these parameters.

\section{The Bose-condensed gas in a momentum space}

We now investigate the density distribution of the Bose-condensed gas
in a momentum space. Note that for the parameters used here, the
Thomas-Fermi approximation is no more valid in the $x$ and $y$ directions.
For a larger value of $U_0$ it is reasonable to assume that the density
distribution of subcondensate in the optical lattice
has a form of Gaussian distribution in $x$-$y$ coordinate space, i.e. $\Phi
_{0}\left( x,y\right) \sim \sum_{k_{x},k_{y}}^{^{\prime }}\exp \lbrack
-\left( (x-k_{x}d)^{2}-(y-k_{y}d)^{2}\right) /2\sigma ^{2}\rbrack $. Due to
the periodicity of the lattice, the wave function in momentum space
takes the form

\begin{equation}
\Psi _{0}\left( p_{x},p_{y}\right) =\Phi _{0}\left( p_{x},p_{y}\right) \frac{%
\sin \left[ \left( 2k_{M}+1\right) p_{x}d/\hbar \right] }{\sin p_{x}d/2\hbar
}\frac{\sin \left[ \left( 2k_{M}+1\right) p_{y}d/\hbar \right] }{\sin
p_{y}d/2\hbar },  \label{momentum}
\end{equation}
where $\Phi _{0}\left( p_{x},p_{y}\right) \sim \exp \lbrack
-(p_{x}^{2}+p_{y}^{2})\sigma ^{2}/2\hbar ^{2}\rbrack $. We see that the
momentum distribution of the Bose-condensed gas exhibits sharp peaks at $%
p_{j}=2\pi n_{j}\hbar /d$ ($j=x,y$). The width of each peak in momentum
space can be approximated as $\Delta p_{j}\sim \pi \hbar /\left(
2k_{M}+1\right) d$. From this character one can obtain directly the following
results:

(i) As pointed
out in \cite{PEDRI,XIONG}, the density distribution of the subcondensates in the
momentum space provides  important information for the evolution of
interference patterns.
The sharp peaks of the momentum distribution of the condensate
with nonzero $n_{j}$ implies
that there would be side peaks in coordinate space after the optical lattice is
switched off. The classical approximation of the initial
velocity $v_{j}=p_{j}/m$ can give a quite well description for the motion of
the side peaks. When only the optical lattice is switched off, due to the
presence of the magnetic trap the harmonic motion of the center of mass of
the side peaks is given by

\begin{equation}
{\bf r}_{n_x,n_y}=\frac{2\pi n_{x}\hbar }{m\omega _{\perp }d}\sin \left( \omega
_{\perp }t\right) {\bf e}_{x}+\frac{2\pi n_{y}\hbar }{m\omega _{\perp }d}%
\sin \left( \omega _{\perp }t\right) {\bf e}_{y},  \label{motion}
\end{equation}
where ${\bf e}_{x}$ and ${\bf e}_{y}$ are two unit vectors along the  $x$ and $%
y$ directions. When both the optical lattices and magnetic trap are
switched off, the motion of the side peaks is given by

\begin{equation}
{\bf r}_{n_x,n_y}=\frac{2\pi n_{x}\hbar t}{md}{\bf e}_{x}+\frac{2\pi n_{y}\hbar t}{%
md}{\bf e}_{y}.  \label{motionboth}
\end{equation}

(ii) From the density distribution in momentum space
$n\left( p_{x},p_{y}\right) =\left| \Phi_{0}\left( p_{x},p_{y}\right) \right| ^{2}$,
the relative
population of the side peaks with respect to the central peak can be
approximated by $\exp \lbrack -4\pi ^{2}(n_{x}^{2}+n_{y}^{2})\sigma
^{2}/d^{2}\rbrack $. This formula is very useful when experimental
parameters are chosen to observe clearly the side peaks in the interference
patterns.

\section{Density distribution of the condensate after switching off only the
optical lattice}

In \cite{XIONG}, evolution of interference patterns of the
subcondensates induced by a 1D optical lattice is investigated when only
the optical lattice is switched off. In this section, we generalize the method
developed in \cite{XIONG} to consider
the evolution of interference pattern when swithching off only the 2D optical
lattice. Using the density distribution in coordinate space
and the result given by Eq. (\ref{ratio}), the normalized wave function of
the Bose-condensed gas at time $t=0$ takes the form

\begin{equation}
\Psi \left( x,y,z,t=0\right) =A_{n}\sum_{k_{x},k_{y}}^{^{\prime }}\Phi
_{k_{x}k_{y}}\left( x,y,t=0\right) \Phi _{k_{x}k_{y}}\left( z,t=0\right) ,
\label{wavet0}
\end{equation}
$\vspace{1pt}\vspace{1pt}$where $A_{n}=\sqrt{5/2}/\pi \sigma k_{M}$ is a
normalization constant. In the above equation, $\Phi _{k_{x}k_{y}}\left(
z,t=0\right) $ represents the normalized wave function in the $z-$direction,
and $\Phi _{k_{x}k_{y}}\left( x,y,t=0\right) $ is given by

\begin{equation}
\Phi _{k_{x}k_{y}}\left( x,y,t=0\right) =\left( 1-\frac{k_{x}^{2}+k_{y}^{2}}{%
k_{M}^{2}}\right) ^{3/4}e^{-((x-k_{x}d)^{2}+(y-k_{y}d)^{2})/2\sigma ^{2}}.
\label{wavexy0}
\end{equation}
Note that the factor $\left( 1-\left( k_{x}^{2}+k_{y}^{2}\right)
/k_{M}^{2}\right) ^{3/4}$ in the above expression reflects the non-uniform
distribution of the atoms in the lattice  which is given by Eq. (\ref
{ratio}).

When the optical lattice is switched off, the evolution of a
interference pattern can be illustrated by considering the density
distribution $n(x,y,t)$ in $x$-$y$ coordinate space:

\begin{equation}
n(x,y,t)=N_{0}\int \left| \Psi \left( x,y,z,t\right) \right|
^{2}dz=N_{0}\left| \Psi \left( x,y,t\right) \right| ^{2}.
\end{equation}

In the case of the effective chemical potential $\mu _{k_{x}k_{y}}$ being
much smaller than the ground state energy $\hbar \widetilde{\omega }_{\perp
}/2$ of the atoms in the lattice site $(k_x,k_y)$,
the noninteracting model can provides a good
description\cite{PEDRI,XIONG}. When neglecting the interatomic interaction,
the normalized wave function $\Psi \left( x,y,t\right)
=A_{n}\sum_{k_{x},k_{y}}^{^{\prime }}\Phi _{k_{x}k_{y}}\left( x,y,t\right) $
can be obtained through the well-known propagator method\cite{XIONG}.
Recently, the vortex dynamics of parabolically confined Bose gases is also
investigated by the propagator method \cite{TEMPERE}. Using this technique the
condensed state wave function at time $t$ can be
obtained by using the following integral equation \cite{FEYNMAN}:

\begin{equation}
\Psi \left( x,y,t\right) =\int_{-\infty }^{\infty
}K(x,y,t;x_{1},y_{1},t=0)\Psi \left( x_{1},y_{1},t=0\right) dx_{1}dy_{1},
\label{wavettt}
\end{equation}
where the propagator $K(x,y,t;x_{1},y_{1},t=0)$ is given by

\begin{equation}
K(x,y,t;x_{1},y_{1},t=0)=\prod_{j=x,y}K_{j}(r_{j},t;r_{j1},t=0).
\label{propagator}
\end{equation}
In the above equation, $r_{x}\equiv x$ ($r_{y}\equiv y$), $r_{x1}\equiv
x_{1} $ ($r_{y1}\equiv y_{1}$), and

\begin{equation}
K_{j}(r_{j},t;r_{j1},t=0)=\left[ \frac{m\omega _{\perp }}{2\pi i\hbar \sin
\omega _{\perp }t}\right] ^{1/2}\exp \left\{ \frac{im\omega _{\perp }}{%
2\hbar \sin \omega _{\perp }t}\left[ \left( r_{j}^{2}+r_{j1}^{2}\right) \cos
\omega _{\perp }t-2r_{j}r_{j1}\right] \right\} .  \label{propagator-harmonic}
\end{equation}

By Eqs. (\ref{wavexy0}) and (\ref{wavettt}) it is straightforward to get
the following analytical result of $\Psi \left( x,y,t\right) $:

\begin{equation}
\Psi \left( x,y,t\right) =A_{n}\sum_{k_{x},k_{y}}^{^{\prime }}\left( 1-\frac{%
k_{x}^{2}+k_{y}^{2}}{k_{M}^{2}}\right) ^{3/4}\prod_{j=x,y}\Xi _{j}\left(
r_{j},t\right) ,  \label{wavefunttt}
\end{equation}
where

\[
{\ }\Xi _{j}\left( r_{j},t\right) {=\sqrt{\frac{1}{\sin \omega _{\perp
}t\left( {\rm cot}\omega _{\perp }t+i\gamma \right) }}\exp \left[ -\frac{%
\left( k_{j}d\cos \omega _{\perp }t-r_{j}\right) ^{2}}{2\sigma
^{2}\sin ^{2}\omega _{\perp }t\left( {\rm cot}^{2}\omega _{\perp
}t+\gamma ^{2}\right) }\right] \times }
\]

\begin{equation}
\exp \left[ -\frac{i\left( k_{j}d\cos \omega _{\perp }t-r_{j}\right) ^{2}%
{\rm cot}\omega _{\perp }t}{2\gamma \sigma ^{2}\sin ^{2}\omega
_{\perp }t\left( {\rm cot}^{2}\omega _{\perp }t+\gamma ^{2}\right)
}\right] \exp \left[ \frac{i\left( r_{j}^{2}\cos \omega _{\perp
}t+k_{j}^{2}d^{2}\cos \omega _{\perp }t-2k_{j}r_{j}d\right)
}{2\gamma \sigma ^{2}\sin \omega _{\perp }t}\right],  \label{xixi}
\end{equation}
where the dimensionless parameter $\gamma =\hbar /m\omega
_{\perp }\sigma ^{2}$.

From the result given by Eq. (\ref{wavefunttt}), we see that the period of $%
\Psi \left( x,y,t\right) $ is $T=2\pi /\omega _{\perp }$, while the period
of the density distribution $n(x,y,t)$ is $\pi /\omega _{\perp }$.
The density at $x=0$, $y=0$ reaches a maximum value at
time $t_{m}=\left( 2m-1\right) \pi /\omega _{\perp }$ with $m$ a positive
integer. At time $t_{m}$, $\left| \Psi \left( x=0,y=0,t_{m}\right) \right|
^{2}$ is given by

\begin{equation}
\left| \Psi \left( x=0,y=0,t_{m}\right) \right| ^{2}=\alpha _{\perp
-ideal}^{2}=\frac{A_{n}^{2}}{\gamma ^{2}}\left( \frac{4\pi k_{M}^{2}}{9}%
\right) ^{2}.  \label{idealdensity}
\end{equation}

According to the fact that there is only a small number of atoms in each lattice
site, we assume here that the atoms in a subcondensate are in the ground
state of the corresponding lattice site. Thus
we have $\sigma =\sqrt{\hbar
/2m\widetilde{\omega }_{\perp }}$ and it is easy to verify that for the
parameters used here satisfy $\sigma <<d$. This shows that in the derivation of
Eq. (\ref{GP2}) the condition $\sigma <<d$ holds. From the result given by
Eqs. (\ref{wavefunttt}) and (\ref{xixi}), one can obtain the density distribution
of the interference pattern. Displayed in figure 1a is the density
distribution $n(x,y,t)$ (in unit of $N_{0}A_{n}^{2}$) at $t=0.3\pi /\omega
_{\perp }$. Note that in all figures plotted in this paper, the coordinates $x$
and $y$ are measured in the unit $d$. The central and side peaks are clearly shown in
the figure. Due to the presence of the magnetic trap, there is a
harmonic motion of the side peaks. Displayed in figures 1b-d are the
central and side peaks of the interference pattern at $t=0.5\pi /\omega
_{\perp }$. From the result shown in figure 1b, we see that the central peak
is a very sharp one, analogously to the case of 1D optical lattice \cite
{XIONG}. Shown in figure 2 is the evolution of the interference pattern at
different times. The evolution in the width of the central and side peaks and
the motion of the side peaks is obvious. Due to the presence of
the magnetic trap,  there is no obvious evolution in the $z$-direction.

\section{Density distribution of the condensate after switching off the combined
potentials}

We now turn to discuss the time evolution of the interference patterns when both
the magnetic trap and the optical lattice are switched off. In this case
the propagator can be obtained by setting
$\omega _{\perp }\rightarrow 0$ in Eq. (\ref{propagator}). Thus we get

\begin{equation}
K_{b}(x,y,t;x_{1},y_{1},t=0)=\prod_{j=x,y}K_{bj}(r_{j},t;r_{j1},t=0),
\label{bpropagator}
\end{equation}
where

\begin{equation}
K_{bj}(r_{j},t;r_{j1},t=0)=\left[ \frac{2\pi i\hbar t}{m}\right] ^{-1/2}\exp
\left\{ \frac{im\left( r_{j}-r_{j1}\right) ^{2}}{2\hbar t}\right\} .
\end{equation}
The analytical result of $\Psi \left( x,y,t\right) $ is then

\begin{equation}
\Psi \left( x,y,t\right) =A_{n}\sum_{k_{x},k_{y}}^{^{\prime }}\left( 1-\frac{%
k_{x}^{2}+k_{y}^{2}}{k_{M}^{2}}\right) ^{3/4}\prod_{j=x,y}\Xi _{bj}\left(
r_{j},t\right) ,  \label{bwavettt}
\end{equation}
where

\begin{equation}
\Xi _{bj}\left( r_{j},t\right) {=}\left( \frac{1}{1+i\Theta }\right)
^{1/2}\exp \left[ -\frac{\left( r_{j}-k_{j}d\right) ^{2}}{2\sigma ^{2}\left(
1+\Theta ^{2}\right) }\right] \exp \left[ \frac{i\left( r_{j}-k_{j}d\right)
^{2}}{2\sigma ^{2}\left( \Theta +1/\Theta \right) }\right],
\end{equation}
where the dimensionless parameter $\Theta =\hbar t/m\sigma
^{2}$. Shown in figure 3a is the density distribution (in unit of $%
N_{0}A_{n}^{2}$) at $t=0.3\pi /\omega _{\perp }$. The density
distribution at $t=0.5\pi /\omega _{\perp }$ is given in figure 3b. We see
that the peaks are wider than those in the case when only the optical lattice is
switched off. Displayed in figure 4 is the evolution of the interference
pattern with the development of time.

As for the evolution in the $z$-direction, analogously to the case of the 1D
optical lattice \cite{PEDRI} one can obtain that the width $R_{z}\left(
t\right) $ of the central peak is described by the
asympotic law $R_{z}\left( t\right) =R_{z}\left( 0\right) \omega _{z}t$
after the central peak is formed.

\section{Discussion and conclusion}

The time evolution of the interference patterns of a
Bose-condensed gas confined in a 2D optical lattice and a magnetic trap
has been investigated by using a
propagator method. Based on this method the analytical results of
the condensed state wave function and motion of the side peaks
have been provided. When the effective
chemical potential in each lattice site is much smaller than
$\hbar \widetilde{\omega }_{\perp }$,
a noninteracting model can give a quite well description for the evolution of
the interference patterns. Nevertheless, the interaction between atoms can
give an important correction to the central peak when only the optical
lattice is switched off. After switching off the optical lattice,
at time $t_{m}$ the central peak is very sharp and hence
the interaction between atoms can not be neglected. Generally speaking,
starting directly from a general mean field theory, one can give the interaction
correction to the density distribution of the central peak. However, this
would be a challenging work due to the fact that there is not a single phase
for the central peak due to the evolution of the wave packet.
In \cite{XIONG}, for a 1D optical lattices a simple method is developed
to investigate the
interaction correction by using a Gaussian density distribution of the
central peak and energy conservation. In the case of the 2D optical lattice,
the analysis is similar and the interaction correction to the maximum
density of the central peak at $t_{m}$ can be calculated straightforwardly.
Assuming that $\beta $ denotes the ratio of the density at $x=0$%
, $y=0$ between the weakly interacting and noninteracting interference
patterns, at  $t_{m}$ we have

\begin{equation}
\beta =\frac{1}{1+E_{int}\left( \alpha _{\perp -ideal}\right) /E_{all}},
\label{interactionco}
\end{equation}
where $\alpha _{\perp -ideal}$ is given in Eq. (\ref{idealdensity}), with

\begin{equation}
E_{all}=N_{0}\left( \frac{\hbar ^{2}}{2m\sigma ^{2}}+\frac{1}{2}m\widetilde{%
\omega }_{\perp }^{2}\sigma ^{2}\right) ,  \label{allenergy}
\end{equation}
and

\begin{equation}
E_{int}\left( \alpha _{\perp -ideal}\right) =\frac{9N_{0}^{2}g\alpha _{\perp
-ideal}^{2}\omega _{z}}{60k_{M}d\omega _{\perp }}.  \label{interenergy}
\end{equation}
For the parameters used here, a simple calculation shows that $\beta =0.43$.
We see that quite differently from the case of 1D optical lattices (see Ref.
\cite{XIONG}), the weak interaction between atoms results in a change
in a deep way for the
density of the central peak. This is not a surprising result due to the fact
that in the case of the 2D optical lattice, the central peak is cigar-shaped,
while the central peak is disk-shaped for 1D optical lattices. Another
possible interaction correction to the central peak at time $t_{m}$ would be
a four-wave mixing which deserves a detailed study in future.

In addition, the interaction between atoms plays an important role in
the colliding process of the side peaks, and this would be an interesting
problem deserving to be explored further. In the case of the Bose-condensed
gas in a 1D
optical lattice, the collisional haloes have been observed in the experiment
\cite{MORSCH}. When a 2D optical lattice is used to confine an  axial
symmetric Bose-condensed gas, there would be a collision between four side
peaks, rather than the collision between two side peaks as in the case of
1D optical lattice. This gives us new opportunity to investigate the
collision between side peaks both theoretically and experimentally. The
method developed here can also be used straightforwardly to study the
cigar-shaped Bose-condensed gas in a 2D optical lattice. For the
cigar-shaped Bose-condensed gas, the harmonic angular frequencies of the
magnetic trap is $\omega _{x}=\omega _{z}>>\omega _{y}$. When the 2D optical
lattice is imposed in $x$ and $y$ directions, after the optical lattice is
switched off the motion of the side peaks is given by

\begin{equation}
{\bf r}_{n_x,n_y}=\frac{2\pi n_{x}\hbar }{m\omega _{\perp }d}\sin \left( \omega
_{x}t\right) {\bf e}_{x}+\frac{2\pi n_{y}\hbar }{m\omega _{\perp }d}\sin
\left( \omega _{y}t\right) {\bf e}_{y}.  \label{cigar}
\end{equation}
We see that the motion of the side peaks in $x$ and $y$ directions is quite
different because $\omega_x>>\omega_y$.
In this case, the side peaks in the $x$-direction has a
collision before that in the $y$-direction, and this gives us an
opportunity for observing the collisions one after another. It is
obvious that the method developed here can also be used to investigate the
interference patterns of the Bose-condensed gases in a 3D optical lattice which has
been considered recently in \cite{BRAZIL} by using a different method.


\section*{Acknowledgments}


This work was supported by the Natural Science Foundation of China under grant
Nos. 10205011 and 10274021.

\newpage

Figure 1: The interference patterns after only the 2D optical lattice is switched
off. Figure 1a shows the density distribution $n\left( x,y,t\right) $ (in the unit
$N_{0}A_{n}^{2}$) at $t=0.3\pi /\omega _{\perp }$. Figures
1b-d show the central (figure 1b) and side peaks (figure 1c for $\left\{
n_{x}=1,n_{y}=0\right\} $ and figure 1d for $\left\{ n_{x}=1,n_{y}=1\right\}
$) at $t=0.5\pi /\omega _{\perp }$. In all figures, the coordinates $x$ and $%
y $ are measured in unit of $d$. Due to the presence of the magnetic trap,
the central and side peaks are very sharp at $t=0.5\pi /\omega _{\perp
}$.

Figure 2: The evolution process of the central and side peaks
after only the 2D optical lattice is switched  off. The
interference patterns are shown for different time-of-flight $t=0$, $0.1\pi
/\omega _{\perp }$, $0.2\pi /\omega _{\perp }$, $0.3\pi /\omega _{\perp }$, $%
0.4\pi /\omega _{\perp }$, and $0.5\pi /\omega _{\perp }$.

Figure 3: The interference patterns after both the 2D optical lattice and the
magnetic trap are switched off.
Figures 3a and 3b show the interference patterns for $t=0.3\pi
/\omega _{\perp }$ and $0.5\pi /\omega _{\perp }$, respectively. The density
distribution $n\left( x,y,t\right) $ is measured by $N_{0}A_{n}^{2}$, while
the coordinates $x$ and $y$ are in the unit of $d$.

Figure 4: The evolution process of the central and side peaks for
for different time-of-flight $t=0$, $0.1\pi /\omega _{\perp }$,
$0.2\pi /\omega _{\perp }$, $0.3\pi /\omega _{\perp }$, $0.4\pi /\omega
_{\perp }$, and $0.5\pi /\omega _{\perp }$
after both the 2D optical lattice and the magnetic trap are switched off.


\end{document}